\documentclass[twocolumn,twoside,slac_two]{revtex4}
\usepackage{graphicx}
\usepackage{fancyhdr}
\begin{document}

\title{Top quark mass: Latest CDF results, Tevatron combination and electroweak
       implications}

%

\author{Costas Vellidis (on behalf of the CDF Collaboration)}
\affiliation{Department of Physics, University of Athens, Athens 15771, Greece
             and Fermi National Accelerator Laboratory, Batavia, IL 60510, USA}

\begin{abstract}
A summary of the most up-to-date top quark mass measurements at CDF is
presented. These analyses use top-antitop candidate events detected in the CDF
experiment at the Tevatron collider with an integrated luminosity of up to
$\sim$3/fb. The combination of all those measurements together with the
corresponding top mass measurements from the concurrently running D0 experiment
at the Tevatron yields a world average of
${\rm M_{t}=[173.1\pm 0.6(stat.)\pm 1.1(syst.)]}$ GeV/c$^{2}$.
\end{abstract}

\maketitle

\thispagestyle{fancy}


\section{Introduction}
The top quark was discovered in 1995 at the Tevatron proton-antiproton collider
at Fermilab by the CDF and D0 collaborations \cite{Abe,Abachi}. The most
intriguing aspect of the top quark is its mass. It is approximately 35 times
the mass of the next most massive fermion, the b quark, and it is very close to
the electroweak scale. Because of its mass, the top quark gives the largest
contribution to loop corrections in the W boson propagator. Within the Standard
Model (SM), the correlation between the top quark mass (${\rm M_{t}}$) and the
W boson mass induced by these corrections allows for setting limits on the mass
of the yet undiscovered Higgs boson, and favor a relatively light Higgs.

According to the SM, at the Tevatron's 1.96 TeV center-of-mass energy top
quarks are predominantly produced in pairs, by ${\rm q\overline{q}}$
annihilation in $\sim$85\% of the cases and by gluon-gluon fusion in the
remaining $\sim$15\% \cite{Cacciari}. Due to its very short life time, which in
the SM is expected to be about 10$^{-25}$ s, the top quark decays before
hadronizing. In the SM the top quark decays into a W boson and a b quark in
almost 100\% of the cases. The W boson can decay either into quarks as a
${\rm q\overline{q}^{\prime}}$ pair which subsequently hadronize or into a
charged lepton-neutrino pair. This allows for a classification of the
${\rm t\overline{t}}$ candidate events into three non-overlapping samples, or
decay channels, which are characterized by different final-state signatures,
branching ratios (BRs), and background contaminations. The {\it all-hadronic}
sample, where both W bosons decay hadronically, is characterized by six or more
jets in the event (about 55\% of the ${\rm t\overline{t}}$ events). The
{\it lepton+jets} sample, where one W decays leptonically and the other
hadronically, is characterized by one electron or muon, four or more jets, and
large missing transverse energy $\not\!{\rm E}_{\rm T}$ in the event (about
38\% of the ${\rm t\overline{t}}$ events). The {\it dilepton} sample, where
both W bosons decay leptonically, is characterized by two leptons, electrons or
muons, two or more jets, and large $\not\!{\rm E}_{\rm T}$ in the event (about
7\% of the ${\rm t\overline{t}}$ events). The lepton+jets sample has the best
compromise between statistics and background contamination. The dilepton sample
is the cleanest at the cost of having the poorest statistics. The background
contamination in all three samples can be greatly suppressed by ``tagging'' the
jets associated with the b quarks. The most common tagging technique is based
on the displacement of the reconstructed jet vertex from the event's primary
vertex due to the relatively long life time of the b-flavored hadrons.

\section{Top quark mass measurements}
The top quark mass is a free parameter in the SM which can be directly measured
at the Tevatron. Top mass measurements have been performed in each channel
using a variety of methods. The best result has been achieved in the
lepton+jets channel, due to its relatively high BR and moderate background.
Recently a boost has been given to the mass accuracy by an innovative technique
which exploits the hadronic products of the W decay in order to constrain the
largest source of systematic uncertainty: the jet energy scale (JES). In this
technique the mass of the two jets from the W decay is required to match the W
mass, allowing for the so called ``JES in situ'' calibration. Thanks to this
technique analyses in the all-hadronic sample have also achieved a better
sensitivity than those in the dilepton channel. Complementary to this
technique, new measurement methods have been recently applied which make use of
only lepton or track-based information in the event and therefore are free of
the JES systematic uncertainty.

Two general methods have been established to measure the top quark mass at the
Tevatron. In the {\bf Template Mehtod (TM)} distributions, or ``templates'', of
variables strongly correlated with the top mass (most typical example is the
event-by-event reconstructed top mass itself) are reconstructed on signal and
background simulated events. In the {\bf Matrix Element Mehtod (ME)} an
event-by-event probability for signal and background is computed as a function
of the top mass (for the signal only) and of the reconstructed observables. The
ME method exploits all of the information in the event by making use of a
leading order ${\rm t\overline{t}}$ production matrix element, convoluted with
parton distribution functions which model the structure of the colliding protons
and transfer functions which are needed to step back from the reconstructed
jets to the hadronizing partons. Both methods use a likelihood to compare data
with the simulated events and extract the top mass. This likelihood is defined
using a combination of signal and background templates (TM) or probabilities
(ME), weighted according to the expected fraction of signal events in the data.

In the next subsections the top quark mass measurements reaching the highest
sensitivity in CDF are described sample by sample. For brevity, not all of the
measurements are reported in this paper.

\subsection{Dilepton}
The dilepton channel is characterized by a final-state signature of two
high-P$_{\rm T}$ charged leptons (electrons or muons), two high-E$_{\rm T}$
b-jets, and large $\not\!{\rm E}_{\rm T}$ from the neutrinos. The largest
amount of background comes from diboson events, Drelll-Yan events, and W+jets
events where one jet fakes a charged-lepton signature. The signal-to-background
(S/B) ratio is relatively high ($\sim$2 without b-tagging). The greatest
challenge in this channel is the impossibility of ``in situ'' JES calibration.
In addition, the kinematics is under-constrained due to the undetected
neutrinos. The ME method deals with this issue by integrating over neutrino
momenta while computing the event probability, whereas the top mass TM needs
some assumptions to constrain the kinematics and reconstruct the event.

\begin{figure}[h]
\centering
\includegraphics[width=80mm]{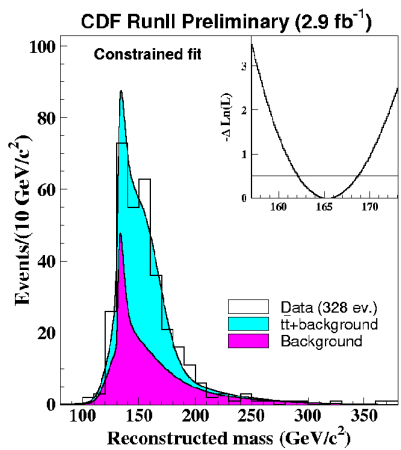}
\caption{The likelihood fit of the neutrino $\phi$ weighting method which
         determines the top quark mass from dilepton events.}
\label{dilepton}
\end{figure}

The most accurate CDF measurement in this channel is based on a ME method
\cite{dilME}. It exploits an evolutionary neural network (NN) optimized
directly on the mass resolution rather than some intermediate or approximate
figure of merit, such as the S/B ratio. The use of a NN improves by 20\% the
mass uncertainty compared to the previous analysis using the same method
\cite{dilMEold}. This measurement yields
${\rm M_{t}=[171.2\pm 2.7(stat.)\pm 2.9(syst.)]}$ GeV/c$^{2}$ for an integrated
luminosity of 2.9/fb. The TM is also used on the basis of an event-by-event top
mass reconstruction \cite{dilTM}. The azimuthal angles of the neutrinos are
integrated in order to constrain the kinematics, hence the method is named
``neutrino $\phi$ weighting''. The likelihood fit, shown in
Figure~\ref{dilepton}, yields
${\rm M_{t}=[165.1^{+3.3}_{-3.2}(stat.)\pm 3.1(syst.)]}$ GeV/c$^{2}$ for an
integrated luminosity of 2.8/fb.

\subsection{Lepton+jets}
The lepton+jets channel is characterized by a signature of a high-P$_{\rm T}$
electron or muon, four high-E$_{\rm T}$ jets, and high $\not\!{\rm E}_{\rm T}$.
The background is mainly composed of W+jets events and multi-jet QCD events in
which a jet is faking the signature of a charged lepton and
$\not\!{\rm E}_{\rm T}$ comes from calorimeter mis-measurements. In order to
enhance the S/B ratio from $\sim$0.5 to $\sim$4 and decrease the possible
jet-to-parton assignments from 12 to 6 the presence of at least one b-tagged
jet is usually required, with an efficiency of $\sim$55\%.

\begin{figure}[h]
\centering
\includegraphics[width=80mm]{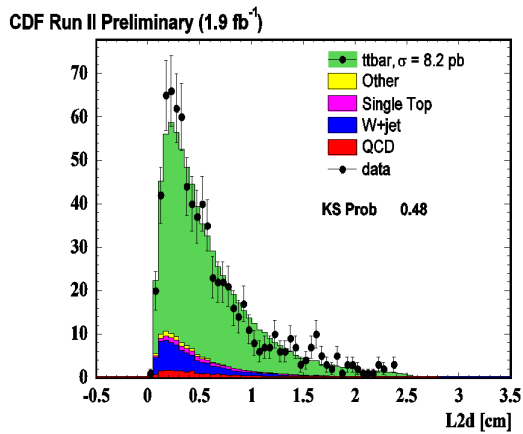}
\caption{The fit of the L$_{\rm 2d}$ signal and background templates which
         determines the top quark mass from lepton+jets events.}
\label{Lxy}
\end{figure}

The most accurate CDF analysis applies the ME method with ``in situ'' JES
calibration \cite{ljME}. The method uses angular and energetic transfer
functions while computing the event probability. The measurement yields
${\rm M_{t}=[172.1\pm 1.1(stat.+JES)\pm 1.1(syst.)]}$ GeV/c$^{2}$ for an
integrated luminosity of 3.2/fb. The b-JES remains the largest source of
systematic uncertainty.

\begin{figure}[h]
\centering
\includegraphics[width=80mm]{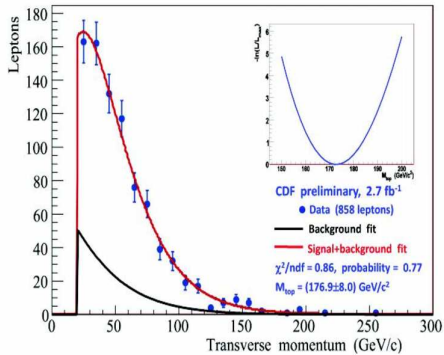}
\caption{The likelihood fit of the lepton P$_{\rm T}$ signal and background
         distributions which determines the top quark mass from lepton+jets
         events.}
\label{leptonPt}
\end{figure}

Two novel TM techniques have been applied to CDF data making no direct use of
jets for measuring the top quark mass. Both make use of kinematic variables
sensitive to the top mass but insenstive to the JES. The one makes use of the
tranverse decay length L$_{\rm xy}$ or L$_{\rm 2d}$ of the b-tagged jets
together with the transverse momentum P$_{\rm T}$ of the leptons and has been
applied to 1.9/fb of lepton+jets data, yielding a result of
${\rm M_{t}=[175.3\pm 6.2(stat.)\pm 3.0(syst.)]}$ GeV/c$^{2}$ \cite{L2d}.
Figure~\ref{Lxy} shows the fit of the L$_{\rm 2d}$ templates to the data. The
other makes use of the transverse momentum P$_{\rm T}$ of the leptons only and
has been applied to 2.8/fb of lepton+jets and dilepton data yielding a combined
result of ${\rm M_{t}=[172.8\pm 7.2(stat.)\pm 2.3(syst.)]}$ GeV/c$^{2}$
\cite{leptonPt}. Figure~\ref{leptonPt} shows the fit of the lepton P$_{\rm T}$
distribution to the data in the lepton+jets channel only. Both techniques are
fast and accurate candidates for the LHC, where the statistics will not limit
the precision of the measurements.

\subsection{All-hadronic}
The all-hadronic channel is characterized by a signature of six
high-E$_{\rm T}$ jets. Current analyses accept events with six to eight jets in
the final state in order to include signal events with additional jets from
initial or final state gluon radiation. At least one b-tagged jet is required.
The all-hadronic sample is challenging because of the huge amount of QCD
multi-jet background. For this reason NN are needed to optimize event selection
in order to drastically enhance the S/B ratio from $\sim$1/400 up to $\sim$1/4.

\begin{figure}[h]
\centering
\includegraphics[width=80mm]{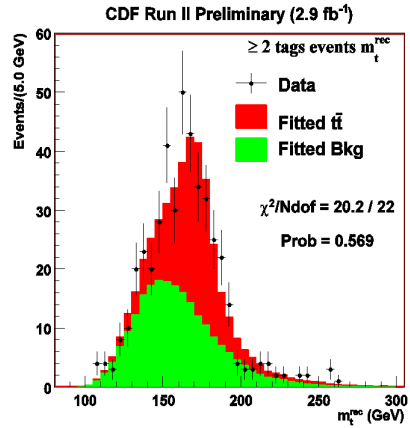}
\caption{The fit of the top mass signal and background templates which
         determines the top quark mass from all-hadronic events.}
\label{alljets}
\end{figure}

So far only CDF has measured the top quark mass from this sample. The most
sensitive analysis in this channel is a 2-dimensional TM \cite{alljets}.
Variables used to build templates are the event-by-event reconstructed top mass
and the JES, which allows for for JES ``in situ'' calibration. This measurement
uses a NN for event selection. This NN was recently upgraded to include also
variables related with the jet shape for a better separation between gluon jets
and light-quark jets from ${\rm t\overline{t}}$ decays. Figure~\ref{alljets}
shows the fit of the top mass signal and background templates which yields a
result of
${\rm M_{t}=[174.8\pm 2.4(stat.+JES)^{+1.2}_{-1.0}(syst.)]}$ GeV/c$^{2}$ with
an integrated luminosity of 2.9/fb.

\section{Tevatron combination, future perspective and electroweak implications}
With the increasing integrated luminosity available at the Tevatron the
systematic uncertainty has started dominating over the statistical uncertainty
in the top quark mass measurements. The JES uncertainty remains the largest one
among the various types of systematics. This is still the case in the
lepton+jets and all-hadronic channels, despite the ``in situ'' JES calibration.

New analyses which follow a different approach to measure the top quark mass
are now emerging in CDF and D0. Such are the b-jet transverse decay length and
the lepton-P$_{\rm T}$ TM anayses described above. Even if these measurements
do not reach a competitive statistical sensitivity, they are reported here
because they are sensitive to different systematic uncertainties compared with
the analyses directly involving jets.

\begin{figure}[h]
\centering
\includegraphics[width=80mm]{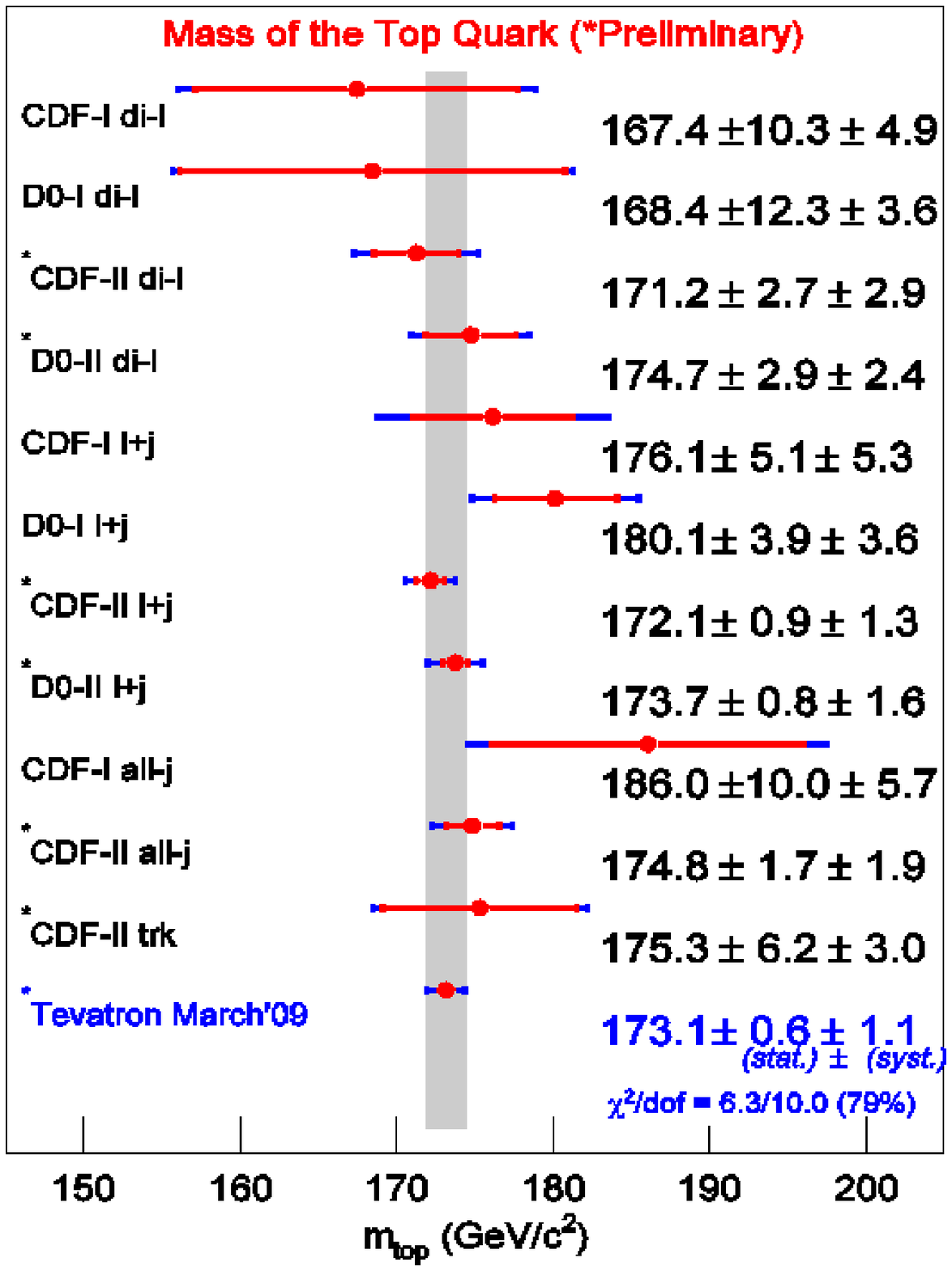}
\caption{Projection of the CDF top quark mass uncertainty as a function of the
         Tevatron integrated luminosity.}
\label{combo}
\end{figure}

Results from most of the analyses discussed above have been used to update
the Tevatron top quark mass combination. Figure~\ref{combo} summarizes the
measurements included in the combination along with the Tevatron combined top
quark mass of ${\rm M_{t}=[173.1\pm 0.6(stat.)\pm 1.1(syst.)]}$ GeV/c$^{2}$, as
of March 2009, which has a relative precision of 0.75\% \cite{TevMt}. CDF by
itself has a top quark mass combination yielding
${\rm M_{t}=[172.6\pm 0.9(stat.)\pm 1.2(syst.)]}$ GeV/c$^{2}$ \cite{CDFMt}. The
precision achieved is below 1\%, already better than the Run II goal.

\begin{figure}[h]
\centering
\includegraphics[width=80mm]{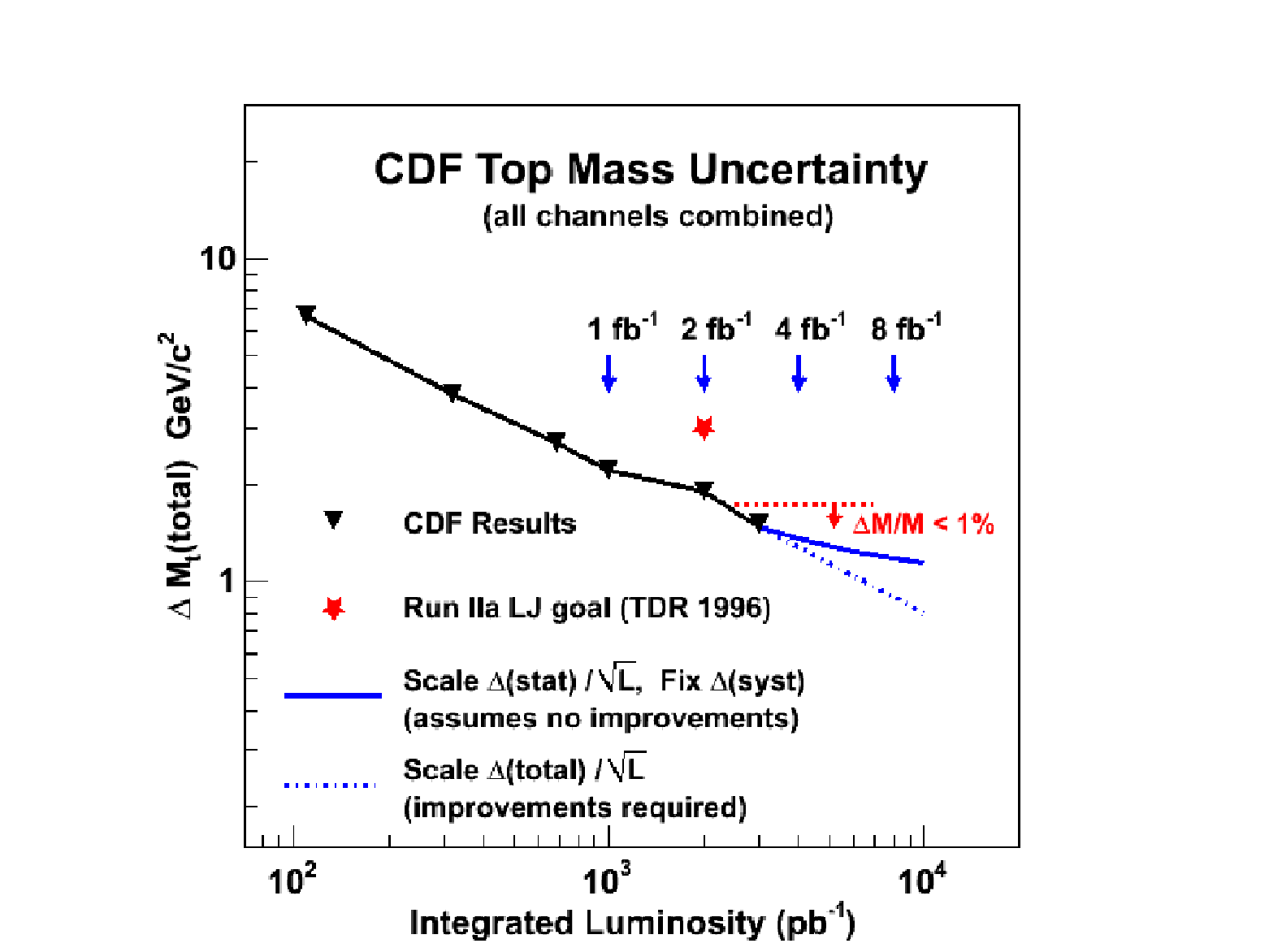}
\caption{Projection of the CDF top quark mass uncertainty as a function of the
         Tevatron integrated luminosity.}
\label{extra}
\end{figure}

The future perspective of CDF for the precision of the top quark mass
measurements is shown in Figure~\ref{extra}. There the top mass total
uncertainty (statistical plus systematic added in quadrature) is shown as a
function of the Tevatron integrated luminosity, with the points representing
the CDF top mass combined results which are obtained so far. The red dashed
line above the last point represents the Run II goal of 1\% relative total
uncertainty. The continuous blue line beyond the last point is an extrapolation
of the total uncertainty assuming that the statistical part will scale with the
luminosity and the systematic part will remain constant, i.e. if no
improvements will be made in the measurement methods. The blue dotted-dashed
line beyond the last point is an extrapolation of the total uncertainty
assuming that both the statistical and systematic parts will scale with the
luminosity. This in turn assumes improvements such that only data driven
sources of uncertainty (e.g. JES calibrations or fakes background estimates)
will dominate the systematic uncertainty.

\begin{figure}[h]
\centering
\includegraphics[width=80mm]{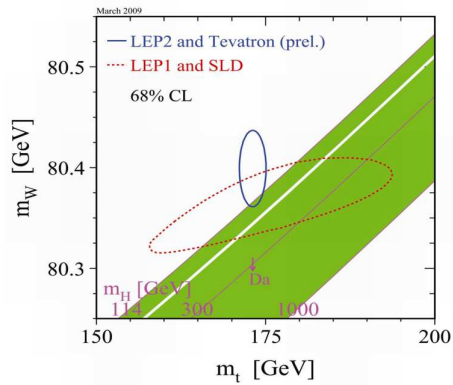}
\caption{1$\sigma$-level expectation for the SM Higgs boson mass derived from
         the measurements of the W boson and top quark masses.}
\label{ewkfit}
\end{figure}

The importance of the high precision of the top quark mass measurements
achieved at the Tevatron for the localization of the SM Higss boson mass, as
discussed in the Introduction, is shown in Figure~\ref{ewkfit}. There the green
band represents regions in the W mass vs. top mass plane corresponding to
different values of the Higgs mass. The ellipsoids represent the expectation
limits set on that plane by the measured W and top masses at the 1$\sigma$
confidence level. The expectation arising from the latest Tevatron top mass
measurement and the combined Tevatron and LEP2 W mass measurements \cite{Wmass}
which is shown by the ellipsoid in continuous blue line points to a low Higgs
mass, as mentioned in the Introduction. The improvement in the localization of
the Higgs mass thanks to the Tevatron top mass precision is shown by comparing
with the old expectation (red dashed line ellipsoid) for which the top mass was
not yet measured but constrained instead by a global electroweak fit.

\begin{acknowledgments}
The author wishes to thank the CDF top-quark working group conveners for their
help in preparing this presentation and the organizers of the ``DPF 2009
Conference'' for their hard work to set up the conference.
\end{acknowledgments}

\bigskip 

\begin{thebibliography}{9}   

\bibitem{Abe} F. Abe et al., (CDF Collaboration), {\it Phys. Rev. Lett.}
              {\bf 74}, 2626 (1995).

\bibitem{Abachi} S. Abachi et al., (D0 Collaboration), {\it Phys. Rev. Lett.}
                 {\bf 74}, 2632 (1995).

\bibitem{Cacciari} S. Abachi et al., JHEP {\bf 04}, 68 (2004); N. Kidonakis and
                   R. Vogt, {\it Phys. Rev. D} {\bf 68}, 114014 (2003).

\bibitem{dilME} A. Abulencia et al., (CDF Collaboration),
                {\it Phys. Rev. Lett.} {\bf 102}, 152001 (2009).

\bibitem{dilMEold} A. Abulencia et al., (CDF Collaboration), {\it Phys. Rev. D}
                   {\bf 75}, 031105 (2007).

\bibitem{dilTM} A. Abulencia et al., (CDF Collaboration), {\it Phys. Rev. D}
                {\bf 79}, 072005 (2008).

\bibitem{ljME} CDF conference note 9692 (2009).

\bibitem{L2d} CDF conference note 9414 (2008).

\bibitem{leptonPt} CDF conference note 9881 (2009).

\bibitem{alljets} CDF conference note 9694 (2009).

\bibitem{TevMt} Tevatron Electroweak Working Group, arXiv:0903.2503 (2009).

\bibitem{CDFMt} CDF conference note 9714 (2009).

\bibitem{Wmass} Tevatron Electroweak Working Group, arXiv:0708.3642 (2007).

\end{thebibliography}

\end{document}